\documentstyle[twocolumn,aps]{revtex}

\input epsf

\begin{document} \title{Measurement of positive and negative scattering lengths in a
Fermi gas of atoms} \author{C. A. Regal and D. S. Jin \cite{adr1}}
\address{JILA, National Institute of Standards and Technology and
Department of Physics, University of Colorado, Boulder, CO
80309-0440} \date{\today} \maketitle

\begin{abstract}  An exotic superfluid phase has been predicted for
an ultracold gas of fermionic atoms.  This phase requires strong
attractive interactions in the gas, or correspondingly atoms with
a large, negative s-wave scattering length.  Here we report on
progress toward realizing this predicted superfluid phase. We
present measurements of both large positive and large negative
scattering lengths in a quantum degenerate Fermi gas of atoms.
Starting with a two-component gas that has been evaporatively
cooled to quantum degeneracy, we create controllable, strong
interactions between the atoms using a magnetic-field Feshbach
resonance.  We then employ a novel rf spectroscopy technique to
directly measure the mean-field interaction energy, which is
proportional to the s-wave scattering length.  Near the peak of
the resonance we observe a saturation of the interaction energy;
it is in this strongly interacting regime that superfluidity is
predicted to occur. We have also observed anisotropic expansion of
the gas, which has recently been suggested as a signature of
superfluidity. However, we find that this can be attributed to a
purely collisional effect.
\end{abstract}
\vskip 0.5 in

\narrowtext

The possibility of superfluidity in an atomic Fermi gas has been
studied
theoretically\cite{Stoof,Holland,Timmermans,Kokkelmans,Hofstetter}.
Analogous to superconductivity in metals and superfluidity in
liquid $^3$He, this fermionic superfluid phase requires
effectively attractive interactions in the gas.  In addition,
these interactions need to be relatively strong if the phase
transition is to occur at a temperature $T_c$ that can be realized
with current experimental techniques. Magnetic-field tunable
Feshbach scattering resonances provide a unique tool that could
create these conditions in an ultracold atomic Fermi gas. These
resonances arise when the collision energy of two free atoms
coincides with that of a quasi-bound molecular
state\cite{Stwalley76,Ties92}.  By varying the strength of the
external magnetic field the experimenter can tune relative
energies through the Zeeman effect and thus control the strength
of the interactions as well as whether they are effectively
repulsive or attractive.

A novel superfluid phase, termed ``resonance superfluidity," is
predicted for a Fermi gas near a Feshbach resonance peak
\cite{Holland,Timmermans}. Unique properties of this phase include
a $T_c$ that is a relatively large fraction of the Fermi
temperature $T_F$.  With predicted $T_c/T_F$ as high as 0.5,
resonance superfluidity may have more in common with exotic
high-$T_c$ superconductors than with ordinary superfluidity. In
fact, resonance superfluidity would exist in a crossover regime
between the physics of Bose-Einstein condensation, which occurs
for composite bosons made up of tightly bound fermions, and
BCS-type superconductivity, which occurs for pairs of fermions
that are loosely correlated in momentum space (Cooper pairs). The
experimental investigation of resonance superfluidity, and its
dependence on magnetic-field detuning from the Feshbach resonance,
could provide new insight into these two types of quantum fluids.

Interactions in ultracold gases arise predominantly from binary
s-wave collisions whose strength can be characterized by a single
parameter, the s-wave scattering length $a$. Across a Feshbach
resonance $a$ can in principle be varied from $-\infty$ to
$+\infty$, where $a<0$ ($a>0$) corresponds to effectively
attractive (repulsive) interactions.  In atomic Fermi gases
Feshbach resonances have been seen in their enhancement of
elastic\cite{Loftus,Regal,O'Hara} and
inelastic\cite{Regal,O'Hara,Jochim,Dieckmann} collision rates.
However collision cross sections depend on $a^2$ and are not
sensitive to whether the interactions are attractive or repulsive.
The mean-field energy, on the other hand, is a quantum mechanical,
many-body effect that is proportional to $na$, where $n$ is the
number density. For Bose-Einstein condensates with repulsive
interactions the mean-field energy and therefore $a$ can be
determined from the size of the trapped
condensate\cite{Inouye,Cornish}, while attractive interactions
cause large condensates to become mechanically
unstable\cite{Gerton,Roberts2}. For an atomic Fermi gas the
mean-field interaction energy has a much smaller impact on the
thermodynamics.  Fermionic atoms obey the Pauli exclusion
principle and occupy higher energy states of the external trapping
potential. Correspondingly a trapped Fermi gas has a lower density
and larger kinetic energy. The mean-field energy thus does not
have an observable effect on the cloud size or momentum
distribution\cite{demarco2001}. In this work we introduce a novel
rf spectroscopic technique that measures the mean-field energy
directly. With this technique we have measured both positive and
negative scattering lengths in a Fermi gas of atoms that is tuned
near the peak of a Feshbach resonance.

\begin{figure} \begin{center} \epsfxsize=2
truein \epsfbox{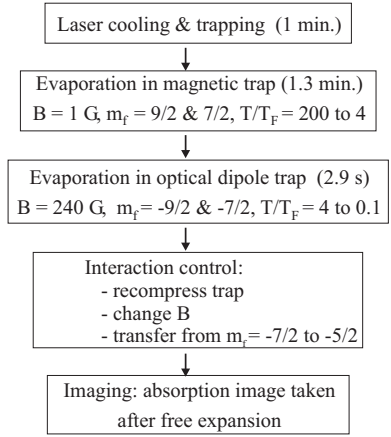} \end{center} \caption{Flowchart of
the experiment.  We start from a room temperature vapor and cool
$10^5$ fermionic atoms to 20 nK in roughly 2.3 minutes. Data are
acquired by repeating this experiment cycle.} \label{flowchart}
\end{figure}

\textbf{Experimental techniques} The experiments reported here
(see Fig. \ref{flowchart}) employ previously developed techniques
for cooling and spin state manipulation of
$^{40}$K\cite{Loftus,DeMarco1}. Because of the quantum statistics
of fermions a mixture of two components, for example atoms in
different internal spin states, is required to have s-wave
interactions in the ultracold gas\cite{DeMarco3}. With a total
atomic spin $f$=9/2 in its lowest hyperfine ground state,$^{40}$K
has ten available Zeeman spin-states $|f, m_f\rangle$.   Atoms in
a 90/10 mixture of the $m_f$=9/2 and $m_f$=7/2 states are held in
a magnetic trap and cooled by forced evaporation \cite{DeMarco1}.
The gas is then loaded into a far-off resonance optical dipole
trap where rf transitions are used to obtain the desired spin
composition \cite{Loftus}. First, the gas is completely
transferred to the $m_f$=-9/2 and $m_f$=-7/2 states using
adiabatic rapid passage. We apply a 15 ms rf frequency sweep
across all ten spin states at a field of $20$ G. We then move to a
higher field, $240$ G, where the Zeeman transitions are no longer
degenerate and apply an rf $\pi/2$ pulse on the $m_f$=-9/2 to
$m_f$=-7/2 transition. After the coherence is lost (which takes
less than 100 ms) we are left with a mixture of two components,
whose relative population can be controlled with the rf pulse
duration.

\begin{figure}
\begin{center} \epsfxsize=3
truein \epsfbox{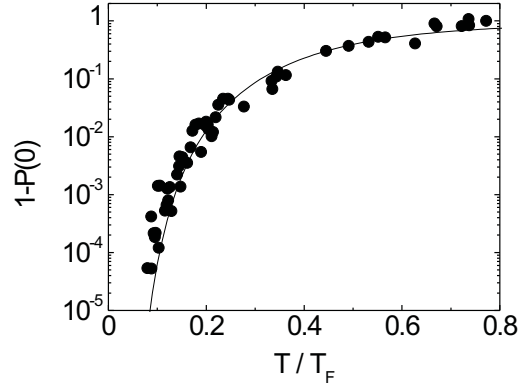} \end{center}\caption{Quantum degeneracy
seen in the thermodynamics of a two-component Fermi gas. Images of
the gas, taken after varying amounts of evaporative cooling, are
fit to a Thomas-Fermi profile with the fugacity $z$ left as a free
fitting parameter that reflects the shape of the cloud. We plot
$1-P(0)=\frac{1}{1+z}$, where $P(0)$ is the occupancy of the
lowest energy state of the trap. The agreement of the data with
the expected curve for an ideal Fermi gas (solid line) confirms
that the gas can be cooled to $T/T_F=0.1$. This agreement is well
within the estimated $18\%$ systematic uncertainty in $T/T_F$ from
the determinations of $N$ and $\nu_z$. The data shown here were
taken with an equal mixture of atoms in the $m_f$=-9/2 and
$m_f$=-7/2 Zeeman states at $B$=240 G.} \label{fugacity}
\end{figure}

With this spin mixture we further evaporatively cool the gas by
lowering the optical trap depth. The optical power is ramped from
900 mW to 8 mW in 2.9 seconds.  During this evaporation, the
radial frequency of the optical trap $\nu_r$ varies from 2800 to
270 Hz and the trap aspect ratio $\lambda=\nu_z/\nu_r$ remains
constant at $0.016$ \cite{freqz}. Starting with 2x10$^7$ atoms at
$T/T_F\approx4$ our evaporation reaches $T/T_F=0.1$ with 10$^5$
atoms remaining. To measure the quantum degeneracy of the
evaporatively cooled gas we take resonant absorption images of the
cloud after expansion from the optical trap.  These images reflect
the momentum distribution of the ultracold gas and are surface fit
to a Thomas-Fermi profile\cite{rokshar}.  In the fits the fugacity
$z$ is left as a free parameter that measures the amount of
distortion from a Gaussian shape.  The value of $z$ obtained from
the fit can then be compared to $T/T_F$ obtained from the measured
temperature $T$ and calculated Fermi temperature
$T_F=h\nu_r(6\lambda N_9)^{1/3}$, where $N_9$ is the total number
of $m_f$=-9/2 atoms and $h$ is Planck's constant. The results of
this analysis, shown in Fig. \ref{fugacity}, verify that the gas
is cooled well into the quantum degenerate regime with $T/T_F$
reaching as low as 0.1.

\begin{figure}  \begin{center} \epsfxsize=2.5  truein
\epsfbox{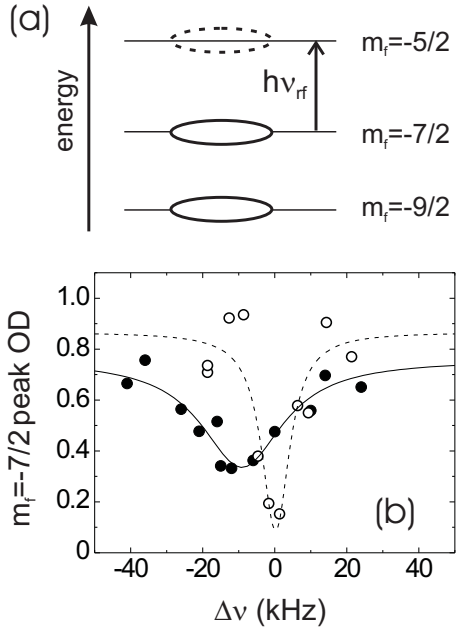} \end{center}\caption{Schematic of the rf
spectroscopy measurement (a) and example rf lineshapes (b). With
atoms originally in the $m_f$=-9/2 and $m_f$=-7/2 Zeeman states,
we apply an rf pulse to transfer atoms from $m_f$=-7/2 to
$m_f$=-5/2. In (b) we plot the peak optical depth (OD) of the
remaining $m_f$=-7/2 gas, measured after expansion, versus rf
frequency $\Delta\nu\equiv\nu_{rf}-\nu(B)$, where $\nu(B)\sim50$
MHz is the expected resonance frequency for a given $B$. The rf
lineshape near the Feshbach resonance ($B=224.36$ gauss)
($\bullet$) is shifted and broadened compared to the lineshape
taken further away from the resonance ($B=215.45$ gauss)
($\circ$). } \label{expt}
\end{figure}

\textbf{Controlling interactions} To control the interactions in
the two-component Fermi gas we access a magnetic-field Feshbach
resonance in the collisions between atoms in the $|9/2,
-9/2\rangle$ and $|9/2, -5/2\rangle$ spin states\cite{newresnote}.
At magnetic fields $B$ near the resonance peak, the mean-field
energy in the Fermi gas was measured using rf spectroscopy (see
Fig. \ref{expt}a).  First, optically trapped atoms were
evaporatively cooled in a 72/28 mixture of the $m_f$=-9/2 and
$m_f$=-7/2 spin states.  After the evaporation the optical trap
was recompressed to either $\nu_r$ = 1390 Hz or $\nu_r$=2770 Hz.
In addition the magnetic field was ramped to the desired value
near the resonance. We then quickly turned on the resonant
interaction by transferring atoms from the $m_f$=-7/2 state to the
$m_f$=-5/2 state with a 73 $\mu$s rf $\pi$-pulse. The fraction of
$m_f$=-7/2 atoms remaining after the pulse was measured as a
function of the rf frequency.

The relative number of $m_f$=-7/2 atoms was obtained from a
resonant absorption image of the gas taken after 1 ms of expansion
from the optical trap. Atoms in the $m_f$=-7/2 state were probed
selectively by leaving the magnetic field high and taking
advantage of nonlinear Zeeman shifts. Sample rf absorption
spectra, taken in the $\nu_r$ = 1390 Hz trap, are shown in Fig.
\ref{expt}b. At magnetic fields well away from the Feshbach
resonance we are able to transfer all of the $m_f$=-7/2 atoms to
the $m_f$=-5/2 state and the rf lineshape has a width limited by
the pulse duration.  The rf frequency for maximum transfer depends
predictably on the current in the magnetic field coils and
provides a calibration of $B$. We estimate that the values of $B$
reported here have a systematic uncertainty of $\pm 0.05$ gauss.
At the Feshbach resonance we observe two changes to the rf
spectra. First, the frequency for maximum transfer is shifted
relative to the expected value from the magnetic field
calibration.  Second, the maximum transfer is reduced and the
measured lineshape is wider.

Both of these effects arise from the mean-field energy due to
strong interactions between $|9/2,-9/2\rangle$ and $|9/2,
-5/2\rangle$ atoms at the Feshbach resonance.  The mean-field
energy produces a density-dependent frequency shift given by
\begin{eqnarray}
\Delta\nu=\frac{2\hbar}{m}n_{9}(a_{59}-a_{79}),
\end{eqnarray}
where $m$ is the atom mass, $\hbar=h/2\pi$, $n_{9}$ is the number
density of atoms in the $m_f=-9/2$ state, and $a_{59}$ ($a_{79}$)
is the scattering length for collisions between atoms in the
$m_f$=-9/2 and $m_f$=-5/2 ($m_f$=-7/2) states. Here we have
ignored a non-resonant interaction term proportional to the
population difference between the $m_f$=-7/2 and $m_f$=-5/2
states; this term equals 0 for a perfect $\pi$-pulse.  For our
spatially inhomogeneous trapped gas, the density dependence
broadens the lineshape and lowers the maximum transfer. This
effect on both sides of the Feshbach resonance peak.  In contrast,
the frequency shift for maximum transfer reflects the scattering
length and changes sign across the resonance.

We have measured the mean-field shift $\Delta\nu$ as a function of
$B$ near the Feshbach resonance peak. The rf frequency for maximum
transfer was obtained from Lorentzian fits to spectra like those
shown in Fig. \ref{expt}b. The expected resonance frequency was
then subtracted to yield $\Delta\nu$.  The scattering length
$a_{59}$ was obtained using Eqn.1 with $n_9=0.5n_p$ and
$a_{79}=174a_o$ where $a_o$ is the Bohr radius\cite{Loftus}. The
peak density of the trapped $m_f$=-9/2 gas $n_p$ was obtained from
Gaussian fits to absorption images. The numerical factor 0.5
multiplying $n_p$ was determined by modelling the transfer with a
pulse-width limited Lorentzian integrated over a Gaussian density
profile in the two radial directions.

\begin{figure}[top]  \begin{center} \epsfxsize=3 truein
\epsfbox{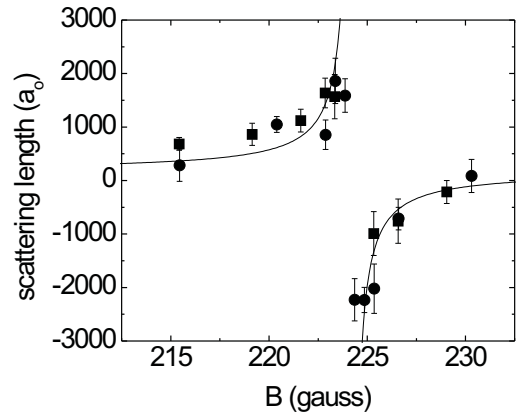} \end{center}\caption{Scattering length versus
magnetic field near the Feshbach resonance peak. The scattering
length $a_{59}$ was obtained from measurements of the mean-field
energy taken for $T/T_F=0.4$ and two different densities:
$n_p=1.8$x$10^{14}$ cm$^{-3}$ ($\bullet$) and $n_p=5.8$x$10^{13}$
cm$^{-3}$ (squares). We estimate a systematic uncertainty in
$a_{59}$ of $\pm50\%$ due to uncertainty in our measurement of the
trapped gas density.} \label{meanfield}
\end{figure}

The measured scattering length as a function of $B$ is shown in
Fig. \ref{meanfield}. This plot, which combines data taken for two
different trap strengths and gas densities, shows that we are able
to realize both large positive and large negative values of
$a_{59}$ near the Feshbach resonance peak. The solid line in Fig.
\ref{meanfield} shows a fit to the expected form for a Feshbach
resonance $a=a_{bg}(1-\frac{w}{B-B_{pk}})$ \cite{Kokkelmans}. Data
within $\pm0.5$ gauss of the peak were excluded from the fit. With
$a_{bg}=174a_o$ we find that the Feshbach resonance peak occurs at
224.21$\pm$0.05 G and has a width $w$ of $9.7\pm0.6$ G.

When $B$ is tuned very close to the Feshbach resonance peak we
expect the measured $a_{59}$ to have a maximum value of
approximately $1/k_F$ due to the unitarity limit. Here $\hbar k_F$
is the Fermi momentum for the $m_f$=-9/2 gas. This saturation can
be seen in the data shown in Fig. \ref{meanfield}. Two points that
were taken within $\pm0.5$ gauss of the Feshbach resonance peak,
one on either side of the resonance, clearly lie below the fit
curve. Observation of this saturation demonstrates that we can
access the strongly interacting regime with $a_{59}k_F$ is greater
than $1$. It is in this regime that resonance superfluidity is
predicted to occur. Using the fit discussed above, we find that
the unitarity-limited point on the attractive interaction side of
the resonance (higher $B$) corresponds to a gas that actually has
$a_{59}k_F\approx11$.

\textbf{Anisotropic expansion} Having obtained a precise map of
$a_{59}$ versus $B$ we investigated the expansion of the Fermi gas
near the Feshbach resonance. Anisotropic expansion has been put
forth as a possible signature of superfluidity\cite{Menotti} and
has recently been observed in a $^6$Li Fermi gas\cite{O'Hara2}.
However, a normal gas in the hydrodynamic regime, where the
collision rate is large compared to the trap oscillator
frequencies, is expected to exhibit the same anisotropy in
expansion\cite{Menotti,Kagan}. In this case, collisions during the
expansion transfer kinetic energy from the elongated axial cloud
dimension into the radial direction.  A large magnitude scattering
length, such as is required for resonance superfluidity, enhances
the collision rate in the gas and could cause the normal gas to
approach the hydrodynamic regime.

We have utilized our ability to tune $a_{59}$ with the
magnetic-field Feshbach resonance to investigate these effects.
After evaporatively cooling atoms in a 45/55 mixture of the
$m_f$=-9/2 and $m_f$=-7/2 spin states, the optical trap power was
increased to yield a trap characterized by $\nu_r$ = 1230 Hz. In
addition, the magnetic field was ramped to the desired value near
the resonance. We then transferred atoms from the $m_f$=-7/2 state
to the $m_f$=-5/2 state with a 29 $\mu$s rf
$\pi$-pulse\cite{pulse}. Expansion was initiated by turning off
the optical trapping beam with an acousto-optic modulator 0.3 ms
after the rf pulse. The magnetic field remained high for 5 ms of
expansion and a resonant absorption image was taken after a total
expansion time of 20 ms.

\begin{figure}  \begin{center} \epsfxsize=2.8 truein
\epsfbox{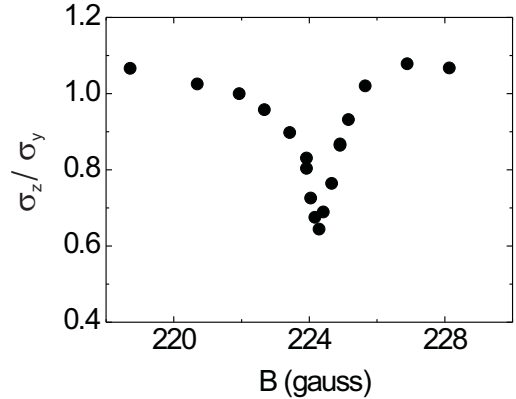} \end{center} \caption{Anisotropic expansion at
the Feshbach resonance peak.  The aspect ratio $\sigma_z/\sigma_y$
of the expanded cloud decreases at the Feshbach resonance because
of an enhanced elastic collision rate. Note that
$\sigma_z/\sigma_y$, measured after 20 ms of expansion, drops
below 1, indicating that the initially smaller radial size has
grown larger than the axial size.  The data were taken for
$N=2.4$x$10^5$ atoms at $T/T_F=0.34$.} \label{anisotropy}
\end{figure}

We find that the ratio of the axial and transverse widths of the
expanded cloud, $\sigma$$_z$/$\sigma$$_y$, decreases at the peak
of the Feshbach resonance (Fig. \ref{anisotropy}).  This effect
depends on having strong interactions during the expansion; we do
not see any change in $\sigma$$_z$/$\sigma$$_y$ if the magnetic
field, and consequently the resonant interactions, is switched off
at the same time as the optical trap. The anisotropic expansion is
basically symmetric about the peak of the resonance and thus is
not sensitive to the sign of the interactions (repulsive or
attractive). Therefore the anisotropic expansion observed here is
not a signature of superfluidity but rather a collisional effect.

To see if hydrodynamic expansion of the normal gas is expected we
can calculate the elastic collision rate $\Gamma$ in the
gas\cite{Holl01}. For this calculation $T$ and $N$ are obtained
from Thomas-Fermi fits to absorption images of clouds away from
the Feshbach resonance where we find that $T/T_F=0.34$. Using an
elastic collision cross section given by $\sigma=4 \pi a_{59}^2$
and $|a_{59}|=2000$ (as was measured near the resonance peak), we
find $\Gamma$=46 kHz. With $\Gamma/\nu_r=37$ and
$\Gamma/\nu_z=2400$ it is not surprising that we observe
anisotropic expansion. For a gas that was fully hydrodynamic, with
$\Gamma\gg\nu_r,\nu_z$, we would expect our measured aspect ratio
to reach 0.4\cite{Menotti,Kagan}.

\textbf{Future prospects} In conclusion, we have measured the
mean-field energy in a Fermi gas of atoms for both strong
attractive and strong repulsive interactions. The rf spectroscopy
technique demonstrated here allows one to measure both the sign
and strength of the interactions and thus take full advantage of
the control over interactions afforded by a magnetic-field
Feshbach resonance.  In addition to the ability to create strong
attractive interactions in the gas, we have demonstrated two other
ingredients that are key to the goal of realizing Cooper pairing
in a gas of atoms.  We are able to cool the optically trapped gas
to $T/T_F\approx0.1$, and precisely control the spin composition.
In addition the rf spectroscopy demonstrated here could provide a
method for detecting ``resonance superfluidity" by measuring the
binding energy of Cooper pairs\cite{Petrosyan,Torma,Bruun}.
Finally, we have observed anisotropic expansion of the strongly
interacting Fermi gas, and find that this is a collisional effect
that occurs for both attractive and repulsive interactions.

\textbf{Acknowledgements}  We thank S. Inouye and M. Holland for
discussions. This work was supported by NSF, ONR, and NIST, and C.
A. R. acknowledges support from the Hertz Foundation.


\begin{thebibliography}{10}
\bibitem{freqz} The radial trap frequency $\nu_r$ was measured by
exciting a shape oscillation of the cloud. The axial frequency
$\nu_z$ was measured by modulating the optical trap power to
parametrically heat the gas.
\bibitem{pulse} The short duration of this pulse gives a frequency
width that allows us to transfer most ($>70\%$) of the $m_f=-7/2$
atoms in spite of mean-field shifts at the Feshbach resonance
peak.
\end{thebibliography}
\end{document}